# Systematic design and experimental demonstration of transmission-type multiplexed acoustic meta-holograms

Yifan Zhu[1†], Nikhil JRK Gerard[2,3†], Xiaoxing Xia[4], Grant C. Stevenson[2], Liyun Cao[1], Shiwang Fan[1], Christopher M. Spadaccini[4], Yun Jing[2,3*], and Badreddine Assouar[1*]

[1]*Institut Jean Lamour, Université de Lorraine, CNRS, Nancy, 54000, France*

[2]*Department of Mechanical and Aerospace Engineering, North Carolina State University, Raleigh, North Carolina 27695, USA*

[3]*Graduate Program in Acoustics, The Pennsylvania State University, University Park, Pennsylvania, 16802, USA*

[4]*Lawrence Livermore National Laboratory, 7000 East Avenue, Livermore, CA94550, USA*

**Abstract**

Acoustic holograms have promising applications in sound-field reconstruction, particle manipulation, ultrasonic haptics and therapy. This paper reports on the theoretical, numerical, and experimental investigation of multiplexed acoustic holograms at both audio and ultrasonic frequencies via a rationally designed transmission-type acoustic metamaterial. The proposed meta-hologram is composed of two Fabry-Pérot resonant channels per unit cell, which enables the simultaneous modulation of the transmitted amplitude and phase at two desired frequencies. In contrast to conventional acoustic metamaterial-based holograms, the design strategy proposed here, provides a new degree of freedom (frequency) that can actively tailor holograms that are otherwise completely passive and hence significantly enhances the information encoded in acoustic metamaterials. To demonstrate the multiplexed acoustic metamaterial, we first show the projection of two different high-quality meta-holograms at 14 kHz and 17 kHz, with the patterns of the letters, N and S. We then demonstrate two-channel ultrasound focusing and annular beams generation for the incident ultrasonic frequencies of 35 kHz and 42.5 kHz. These multiplexed acoustic meta-holograms offer a technical advance to tackle the rising challenges in the fields of acoustic metamaterials, architectural acoustics, and medical ultrasound.

These authors contributed equally: Yifan Zhu, Nikhil JRK Gerard
**Corresponding authors:**
*yqj5201@psu.edu
*badreddine.assouar@univ-lorraine.fr

Drawing inspiration from their optical counterparts, acoustic holograms have been recently reported [1-8], and now offer fresh perspectives on a wide-range of applications like advanced sound-field reconstruction, ultrasonic therapy, haptics, and particle manipulation. In this regard, transducer arrays [2-4], acoustic metamaterials [5], and metasurfaces [6-9] are often frequent candidates for the generation of holographic images that possess unique features and functionalities. Conventional approaches [1-6], however, are largely based on pure phase modulation and therefore necessitate iterative computational optimization, such as the Gerchberg–Saxton algorithm [6]. Despite this, such approaches are still accompanied by sub-optimal image quality since the simultaneous modulation of both amplitude and phase is crucial for the reconstruction of high-precision wave fields.

To address this limitation, some recent studies have demonstrated that metasurfaces equipped with precise control over phase and amplitude could offer a promising route for high-fidelity acoustic holography [7-11]. It was illustrated in a prior study, that reflection-type lossy acoustic metamaterials (LAM) [8] could facilitate decoupled amplitude and phase modulation (APM) and therefore allow for the realization of high-quality images that were previously unattainable. Further, a major advantage is that a straightforward, deterministic time-reversal approach could be employed for the calculation of the required APM profile, and hence bypasses the need for complex computational optimization.

Meanwhile, the past few years have also witnessed the advent of multiplexed optical metasurface-based holograms which are encoded with several predesigned patterns or functionalities, that can be selectively projected on to image planes based on the helicity [12-13], angle [14], frequency [15-16] or the direction [17] of the incident wave. Similarly, our recent paper [18] put forward the concept for a reflection-type multiplexed acoustic metamaterial (RMAM) that enables high-fidelity frequency dependent holograms for audible sound. Each unit cell of the RMAM comprises multiple Helmholtz resonators (HRs) [19-22] that are tuned to distinct operating frequencies. These HRs hence possess the geometrical entities that can be utilized to precisely control the amplitude and phase responses at each of these frequencies independently. Although both LAM and RMAM give rise to advanced routes for acoustic holography, these concepts have remained limited to reflective metamaterials in the audible regime. Multiplexing for transmitted wave meta-hologram, on the contrary, holds great potential for a wider range of acoustic wave-based applications as the interference between the hologram and the incident wave is substantially reduced.

Extending prior designs to transmission-based systems, however, is a challenging problem for both optics and acoustics and has hence rendered all prior frequency multiplexed optical [15-16] and acoustical [18] meta-holograms to be of reflection-type. In acoustics, a single HR is insufficient to induce and tailor the required non-trivial local phase shift for transmitted waves and therefore implies more complicated geometries [23]. These are particularly cumbersome to fabricate for higher operating frequencies, where the feature sizes of the micro-structure are much smaller. Additionally, the amplitude modulation in the LAM and RMAM designs is made possible by artificially introducing loss in a controlled manner via leaky slits and absorbing boundaries, that are unfeasible in the case of transmission. New strategies for transmission-type multiplexed metamaterials are thus highly desirable, in order to advance acoustic holography across a broad range of frequencies.

In this paper, we put forward transmission-type multiplexed acoustic meta-holograms (MAMH) and experimentally illustrate their working for both audible sound and ultrasound. The physical mechanism hinges on unit cells that consist of multiple Fabry-Pérot (FP) resonant channels [24-25], whose effective heights correspond to the distinct operating frequencies of interest. Each of these FP resonant channels in turn consists of amplitude and phase modulating sub-structures that allow for precise and simultaneous APM. Furthermore, the amplitude modulation in this case, is made possible by elegantly engineering the inherent dissipation that exists in the sub-wavelength FP channels owing to the thermal and viscous boundary layer effects [26-28]. This is highly beneficial for ultrasonic applications, where thermo-viscous losses are very high and are at present deemed extremely undesirable [29-30]. We employ a two-channel version of our unit cell to systematically design and illustrate frequency dependent functionalities that leverage simultaneous APM and significantly increase the degrees of freedom for conventional acoustic wave manipulation. To this end, we demonstrate multiplexed high-quality audio holography at 14 kHz and 17 kHz, and two-channel ultrasonic focusing and frequency dependent ultrasonic annular beam generation at 35 kHz and 42.5 kHz. These demonstrations could pave the way for a host of multiplexed and switchable compact devices that could facilitate various applications in room acoustics, audio engineering, virtual reality, and biomedical ultrasound.

**Results**

**Working principle of the MAMH.** Figures 1(a) and (b) illustrate the working principle of the MAMH, where the incident wavefront consists of two frequencies, *viz*., $f_1$ and $f_2$, marked by the green and blue arrows, respectively. For multiplexed transmission holography, when the wave propagates through the MAMH, it must be reshaped such that it projects different predesigned images for $f_1$ and $f_2$, on to an image plane that exists at a given distance in front of it. The desired holographic patterns can be arbitrarily chosen and captured by the acoustic pressure field, denoted here by, $P_n(x, y, z)$, where $n$ corresponds to the operating frequency ($n = 1$ or 2 in this study). In order to reconstruct this image via a meta-hologram, we first employ the time-reversal (TR) method [8] to calculate the pressure profile, $p_n(x_j, y_j)$, that must exist at the hologram plane. We then design the meta-hologram such that it facilitates an amplitude and phase distribution, $(A_{n(TR)}, \varphi_{n(TR)})$, that is consistent with $p_n(x_j, y_j)$ on the hologram plane and therefore satisfies the desired $P_n(x, y, z)$ at the image plane for the $n^{th}$ frequency.

Foremost, $P_n(x, y, z)$ is decomposed into a collection of pixels that are represented in Fig. 1(a), by $p_n(x_l, y_l, z_l)$, where $l$ denotes the pixel number. The pressure at each of the $l^{th}$ pixels can in turn be defined as, $p_n(x_l, y_l, z_l) \equiv p_l \equiv A_{0l}\exp(i\varphi_{0l})$, where $A_{0l}$ and $\varphi_{0l}$ are the predesigned amplitude and phase of a pixel that is located at a point, $(x_l, y_l, z_l)$, on the image plane. In hologram design, we usually set $\varphi_{0l} = 0$, which means that the phase distribution of the predesigned image is uniform. Likewise, the hologram plane can be described by a collection of pixels, and the pressure at every $j^{th}$ pixel can be written as, $p_n(x_j, y_j) \equiv p_j \equiv A_j\exp(i\varphi_j)$ with amplitude $A_j$ and phase $\varphi_j$. Since the system obeys time-reversal symmetry, $A_j$ and $\varphi_j$, can be calculated by simply back projecting the information on the image plane as follows,

$$p_j = \sum_{l=1}^{N} \frac{A_{0l}}{r_l} \exp[i(k_n r_l + \phi_{0l})], \tag{1}$$

where $k_n = 2\pi f_n/c_0$ is the wave number, $c_0 = 343$ m/s is the sound speed in air. $N$ is the total number of image pixels, and the distance, $r_l$, between the image pixel and hologram pixel can be calculated as, $r_l^2 = (x_j-x_l)^2+(y_j-y_l)^2+(z_j-z_l)^2$. Similarly, by employing the values of $p_j$, we can reaffirm its accuracy by analytically calculating the pressure distribution at the image plane through the following formula

$$P_n(x, y, z) = \sum_{j=1}^{M} \frac{A_j}{r_j} \exp[-i(k_n r_j - \phi_j)], \tag{2}$$

where $M$ is the total number of hologram pixels and the distance, $r_j$ between the spatial point at $(x, y, z)$ and the hologram pixel, $(x_j, y_j, z_j)$, satisfies, $r_j^2 = (x-x_j)^2+(y-y_j)^2+(z-z_j)^2$. The calculated values of $A_j$ and $\varphi_j$, obtained from Eq. (1), would hence serve as the required values of $(A_{n(TR)}, \varphi_{n(TR)})$ and can be arranged as a function of $x$ and $y$ in order to realize multiplexed holography.

It is quite evident from the above, that an MAMH must possess structural units that are endowed with complete control over the amplitude and phase of the transmitted wave, at each of these frequencies independently. Therefore, in what follows, we first delineate the design of our unit cell for the target audio frequencies, $f_1$ = 14 kHz, and $f_2$ = 17 kHz, and then couple this with the aforementioned technique to demonstrate high-precision frequency dependent holography at $f_n$. The latter part of the paper will then put forward the extension of this design to the ultrasonic regime and the subsequent realization of two-channel focusing and annular beam generation.

**Unit cell design.** The MAMH is made up of 3D unit cells with multiple resonant sub-units, as shown in Fig. 1(c). Each unit cell has two FP resonant channels [24-25], which are denoted here by $C_1$ and $C_2$ for channels 1 and 2, respectively. The structure of the channels is made by a solid material and the background medium is air. The size of the unit cell is 10 mm×10 mm×45 mm and the transverse dimensions are smaller than half the wavelength for both cases (i.e., $\lambda_1$ = 24.5 mm, $\lambda_2$ = 20.2 mm). It ensures that only the fundamental mode propagates, independent of the angle of incidence. Further, as illustrated in Fig. 1(c), the unit cell comprises frequency, amplitude, and phase modulators, *viz.*, the FM, AM, and PM, that are marked in blue, black, and red, respectively.

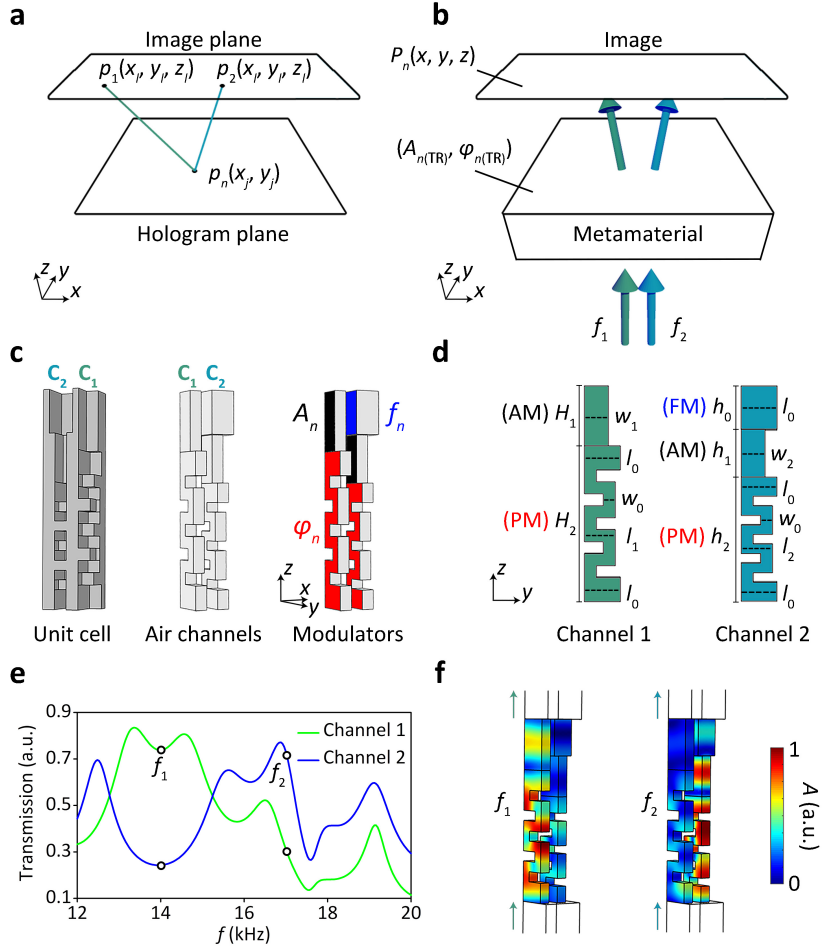

**Fig. 1** The schematic diagram of MAMH. (**a**) The schematic diagram of the hologram plane and image plane. (**b**) The schematic diagram of the MAMH that can generate two different images at two frequencies $f_1$ and $f_2$. (**c**) The unit cells consist of two channels (1 and 2) corresponding to two Fabry–Pérot resonant frequencies, $f_1$ and $f_2$. The figures show the 3D structure of the unit cell and the background medium of the air channel. The schematic diagram of the unit cell consisting of three modulators, *viz.*, the frequency modulator (FM), amplitude modulator (AM) and phase modulator (PM) marked by blue, black and red, respectively. (**d**) The 2D structure of $C_1$ and $C_2$. The parameters are marked in the figure. (**e**) The transmission curve of $C_1$ and $C_2$ from 12 kHz to 20 kHz. At $f_1$=14 kHz (green line), the transmission of $C_1$ is much higher than $C_2$. At $f_2$ = 17 kHz (blue line), the transmission of $C_2$ is much higher than $C_1$. (**f**) The acoustic pressure amplitude fields in the unit cells at $f_1 = 14$ kHz and $f_2 = 17$ kHz, respectively. The arrows indicate the incident and transmitted directions.

Figure 1(d) shows the 2D cross-section of $C_1$ and $C_2$, along with the relevant geometrical parameters. $C_1$ is composed of AM and PM regions with heights $H_1 = 12.5$ mm and $H_2 = 32.5$ mm, respectively. The total height of the two regions, $H$, is therefore the height of the entire unit cell ($H = H_1 + H_2 = 45$ mm). The AM is an air-channel on the top that has a width, $w_1$, which can be tailored to be between 3 mm and 8 mm to predominantly modulate the amplitude of the transmitted wave. The PM is a zig-zag channel whose total effective length can be controlled by tuning the parameter $l_1$ between 3 mm and 8 mm and modulates the transmitted phase. The width of this zig-zag channel, $w_0 = 2.5$ mm and its top and bottom are connecting channels with a fixed width of $l_0 = 7.5$ mm.

To start with, we first design $C_1$ by matching its resonance frequency to the first desired working frequency, $f_1=14$ kHz. This is done by choosing the values of $H_1$ and $H_2$ such that its sum satisfies the FP resonance condition for $f_1$. The $C_2$ is then designed by including a wider air channel, on top of the AM, that matches well with the impedance of the free space. This wider channel is defined as FM, because its height $h_0 = (1-\beta)H$ is changed to choose the second targeted frequency. Here, the $\beta$ is a factor of reduction that is applied to the heights of the AM and PM via $h_1 = \beta H_1$ and $h_2 = \beta H_2$ respectively and allows us to tailor the frequency at which the FP resonance of $C_2$ occurs. In this case, we search the value of $\beta$ and set $\beta=0.8$, so that the heights of the FM, AM and PM are $h_0 = 9$ mm, $h_1 = 10$ mm, and $h_2 = 26$ mm, respectively, and the FP resonance for the $C_2$ (which corresponds to $h_1 + h_2$) occurs at the second desired working frequency, $f_2 = 17$ kHz. It is also useful to note here that, $\beta \approx f_1/f_2$, due to the fact that the resonant wavelength of the acoustic meta-materials can be approximately scaled up with the structural size [31-36]. For more elaborate analytical and descriptive details about the unit cell, the reader is referenced to Supplementary Note 1.

The properties of the conceived sub-units were then assessed by numerically calculating the transmission curves for $C_1$ and $C_2$, in the frequency range between 12 and 20 kHz. These results are shown in Fig. 1(e) and illustrate that at $f_1$, the transmission through $C_1$ (0.74) is much higher than that of $C_2$ (0.24). Likewise, at $f_2$, the transmission of $C_2$ (0.73) is much higher than $C_1$ (0.31). Further, Fig. 1(f) shows the acoustic pressure amplitude field through a unit cell made up of these channels, at $f_1$ and $f_2$, respectively. It can be clearly seen here that when $f_1$ ($f_2$) is incident, the response from $C_1$ ($C_2$) is dominant and that of $C_2$ ($C_1$) results in a very low amplitude background noise. This verifies our theory that the two-channel FP resonance mechanism allows for frequency selective wave transmission with a negligible background noise. Additionally, the transmitted amplitude ($A$) and phase ($\varphi$) for each frequency can be calculated as a function of the different geometrical parameters of $C_1$ and $C_2$, respectively. Figures 2(a) and (b), show the various ($A_1$, $\varphi_1$) and ($A_2$, $\varphi_2$), that can be achieved by modulating the geometrical parameters ($w_1$, $l_1$) and ($w_2$, $l_2$), for $f_1$ and $f_2$, respectively. It is interesting to note here, that the numerical calculations employed take into consideration the thermo-viscous dissipation that exists in these channels and contribute to the amplitude modulation that is achieved. A design library was then built for these working frequencies, with 4 discrete values of amplitude and 8 of phase. The 4 values of amplitude that were chosen are 0.32, 0.48, 0.64, and 0.8 and were normalized to 0.4, 0.6, 0.8, and 1, respectively. The values of phase were discretized between $\pi/4$ and $2\pi$, with a step size of $\pi/4$. These 4 × 8 combinations are shown in Figs. 2(c) and (d) and have ($w_1$, $l_1$) and ($w_2$, $l_2$) dimensions that both lie between 3 mm and 8 mm, with a step size of 0.5 mm between them. Therefore, a MAMH that comprises these unit cells can be conveniently realized via a commercial extrusion-based 3D printer.

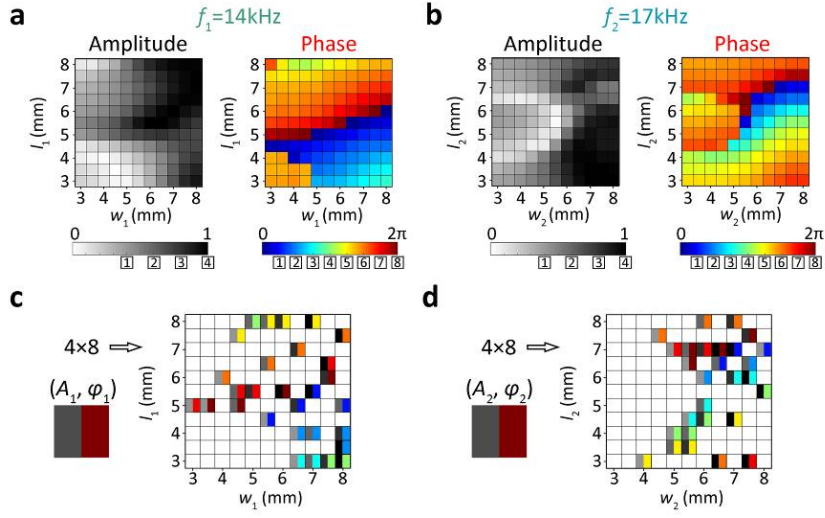

**Fig. 2** Amplitude and phase responses of the unit cell for 14 kHz and 17 kHz. (**a**) The amplitude and phase responses as a function of the parameters $w_1$ and $l_1$ for $f_1$ = 14 kHz. (**b**) The 4×8 combinations of $(A_1, \varphi_1)$ varying with $w_1$ and $l_1$. (**c**) The amplitude and phase responses varying with parameters $w_2$ and $l_2$ for $f_2$ = 17 kHz. (**d**) The 4 × 8 combinations of $(A_2, \varphi_2)$ varying with $w_2$ and $l_2$.

**High-fidelity multiplexed audio holography.** The aforementioned methods were then employed to design a meta-hologram that could selectively project the letters N and S for $f_1$ = 14 kHz and $f_2$ = 17 kHz, respectively, on to an image plane that sits 5 cm away ($z_l$ = 5 cm) from the hologram plane ($z_j$ = 0 cm). The source employed in our experiments was a loudspeaker of diameter 10 cm that was placed 10 cm behind the sample. This was hence regarded as a point source in the full-wave numerical simulations, and the associated amplitude and phase compensation, ($A_{source}$, $\varphi_{source}$) were incorporated in our design to improve its experimental accuracy. This was done by adding a compensation to $A_{n(TR)}$ and $\varphi_{n(TR)}$, to obtain the final amplitude, $A_n = A_{n(TR)}/A_{source}$ and phase, $\varphi_n = \varphi_{n(TR)} - \varphi_{source}$. Figures 3(b) shows the amplitude, $A_{n(TR)}$, and phase, $\varphi_{n(TR)}$, distributions that were calculated by time-reversal for the holographic rendering of the letters, N and S. The plots adjacent to them, $A_n$ and $\varphi_n$, represent the corrected values of the same, upon including the source compensation. The final amplitude and phase profiles that were used for the final MAMH, are shown in Fig. 3(c) as a function of $x$ and $y$.

The design library put forward in the previous section was then revisited to obtain the dimensions for a set of 17 × 17 unit cells that could build the desired MAMH sample of size 18 cm × 18 cm × 4.5 cm (including wall thickness). Figure 3(d) shows the 3D printed prototype of the MAMH that was experimentally characterized by mapping the field over a 20 cm × 20 cm region, at $z = 5$ cm. Figures 4(a) and (b) present the simulated and measured acoustic intensity distributions on the transmissive side of the metamaterial for both frequencies of incidence. The arrows in Fig. 4(a), indicate the direction of propagation and show the numerically calculated intensity field at $z = 4$ cm, 5 cm, and 6 cm away from the sample. The letters 'N' and 'S' are distinctly visible for the two cases and are unimpaired when $z = 5$ cm, proving that the performance of the sample is consistent with our conceived design. Figure 4(b) also shows that the measured acoustic intensity field at $z = 5$ cm, agrees very well with the simulation results and validates the ability of the MAMH to generate high-quality frequency dependent holograms, owing to its simultaneous APM and multiple FP resonance-based mechanism. This unique capability could enable new audio devices and also greatly help renovate existing applications like stereo-sound field reconstruction [37].

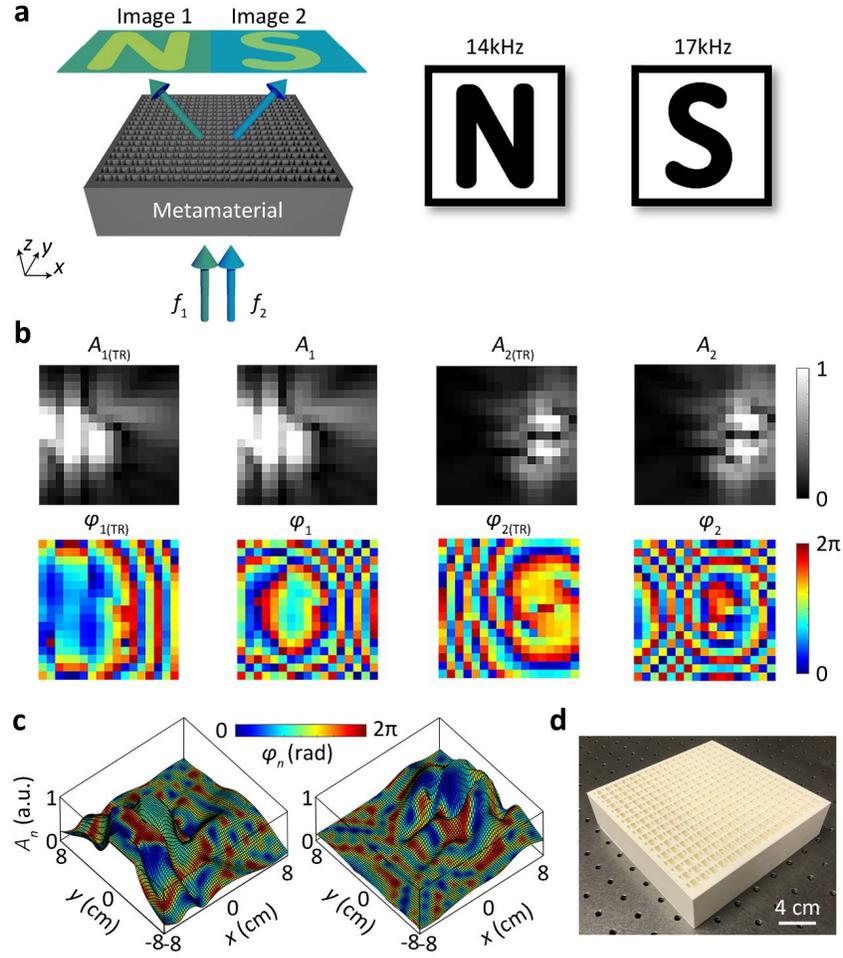

**Fig. 3** The holographic designs for letters "N" and "S" at 14 kHz and 17 kHz, respectively. (**a**) The predesigned holographic image at 14 kHz is letter "N". The image plane is 5 cm away ($z = 5$ cm) from the hologram plane. (**b**) The amplitude and phase distributions at hologram plane calculated by time reversal are ($A_{n(TR)}$, $\varphi_{n(TR)}$). The final amplitude and phase distributions of the sample are calculated as ($A_n$, $\varphi_n$). (**c**) The final distributions of ($A_1$, $\varphi_1$) and ($A_2$, $\varphi_2$). (**d**) The photograph of the sample with a size of 18 cm × 18 cm × 4.5 cm with 17 × 17 unit cells. Scale bar, 4 cm.

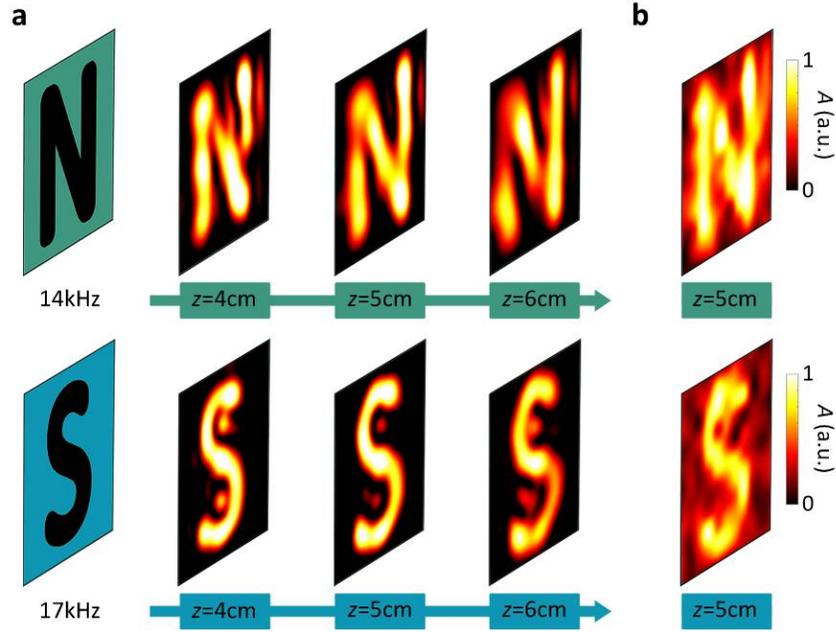

**Fig. 4** Simulated and experimental results of holographic images (letter "N") at 14 kHz and (letter "S") at 17 kHz, respectively. (a) The simulated acoustic intensity distributions at the distance of $z$ = 4 cm, $z$ = 5 cm, $z$ = 6 cm, respectively. The arrows indicated the wave is propagating along the $z$-direction. (**b**) The measured acoustic intensity distributions at the predesigned image plane of $z$ = 5 cm.

**Frequency dependent ultrasonic holography.** The principle and design of our meta-hologram can also be readily extended to operate at ultrasonic frequencies. This is of interest, since multiplexing could greatly benefit ultrasonic applications like therapy and particle manipulation, by circumventing the need for bulky and expensive phased array systems that currently serve as the state of the art for ultrasound beam forming. Besides, the strong intrinsic thermo-viscous loss in ultrasonic metamaterials in fact facilitates the amplitude modulation in our design. Figures 5(a) and (b) show the amplitude and phase responses for our unit cell when it is scaled down by a factor of 0.4. The unit cell now has the dimensions 4 mm × 4 mm× 1.8 mm, and the frequencies of incidence are modified to $f_1$ = 35 kHz and $f_2$ = 42.5 kHz. The design library was built as done previously, with 4 × 8 combinations of $A_1$ and $\varphi_1$ ($A_2$ and $\varphi_2$) with $w_1$ ($w_2$) and $l_1$ ($l_2$) now varying between 1.2 mm and 3.2 mm, with a step size of 0.2 mm, as shown in Figs. 5(c) and (d). This unit cell library was then leveraged to design two multiplexed ultrasonic meta-holograms (MUMH) for the functionalities of frequency dependent focusing and annular beam generation, respectively.

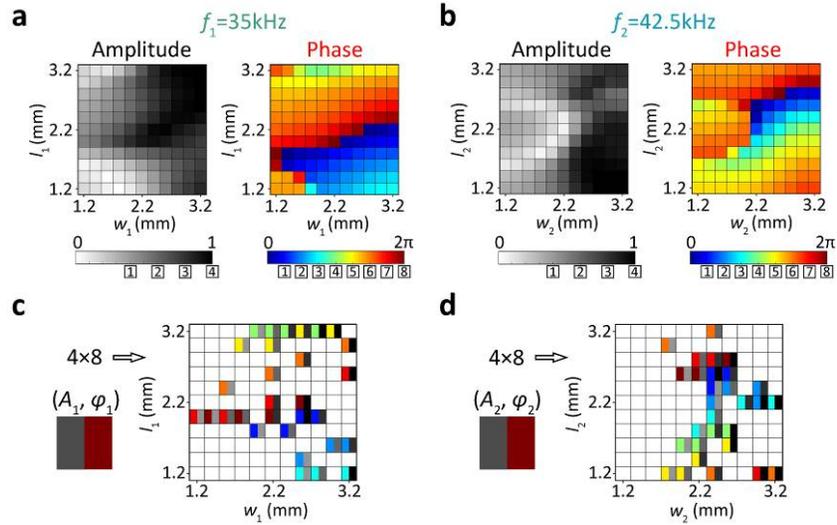

**Fig. 5** Amplitude and phase responses of the unit cell for 35 kHz and 42.5 kHz. (**a**) The amplitude and phase responses changing with parameters $w_1$ and $l_1$ for $f_1$ = 35 kHz. (**b**) The 4 × 8 combinations of ($A_1$, $\varphi_1$) varying with $w_1$ and $l_1$. (**c**) The amplitude and phase responses varying with parameters $w_2$ and $l_2$ for $f_2$ = 42.5 kHz. (**d**) The 4 × 8 combinations of ($A_2$, $\varphi_2$) changing with $w_2$ and $l_2$.

Figure 6(a) shows the schematic for an MUMH that acts as a two-channel ultrasonic lens [38] which exhibits frequency dependent focusing on an image plane that is 1 cm away ($z_l$ = 1 cm). The predesigned holographic images in this case are that of focal points that exist at two different corners of the image plane i.e. (-1.1 cm, 1.1 cm, 1 cm) and (1.1 cm, -1.1 cm, 1 cm), for $f_1$ and $f_2$, respectively. Figure 6(b) here shows the required amplitude and phase distributions that were calculated by the method described earlier. To enable this frequency dependent pressure profile, the MUMH sample was designed with an appropriate set of 10 × 10 unit cells and was fabricated by a commercial high-resolution stereolithography based 3D printer. The resultant prototype of size 4.8 cm × 4.8 cm × 1.8 cm (including wall-thickness) is shown in Fig. 6(c). To experimentally illustrate its working, an air-coupled ultrasonic transducer of diameter 1 cm (center frequency 40 kHz), was placed well-behind the sample (i.e. in the far-field, approx. 5 cm away) such that the wavefront that impinges on the MUMH has a uniform phase distribution and could be captured by a plane wave incidence in the full-wave simulations. The left panel of Fig. 6(d) shows the numerically calculated pressure amplitude distribution on an image plane of size, 4.8 cm × 4.8 cm, situated in front of the sample at $z = 5$ cm. It can be clearly seen here that the designed MUMH enables frequency dependent wave focusing that is consistent with the predesigned holographic images show in Fig. 6(a). This is also evident in our experimentally measured results, shown in the right panel of Fig. 6(d), which agree well with the numerically calculated ones and thereby validates the functionality of our prototype. To further verify the accuracy of the frequency dependent focusing and to compare the simulations with experiments, as shown in Fig. 6(e), the pressure amplitude along the diagonal (described by $x + y = 0$, $z = +1$, shown in the inset) of the image plane was calculated. Here, the $x$ axis denotes the distance between the targeted coordinate and the ordinate (0 cm, 0 cm, 1 cm). The curve calculated from the measurement (squares) has a very reasonable agreement with the numerical results (solid line). Our realization of frequency dependent ultrasonic focusing via a meta-hologram, could greatly aid biomedical applications, since high intensity focused ultrasound is now an emerging candidate for treating different diseases and conditions [39].

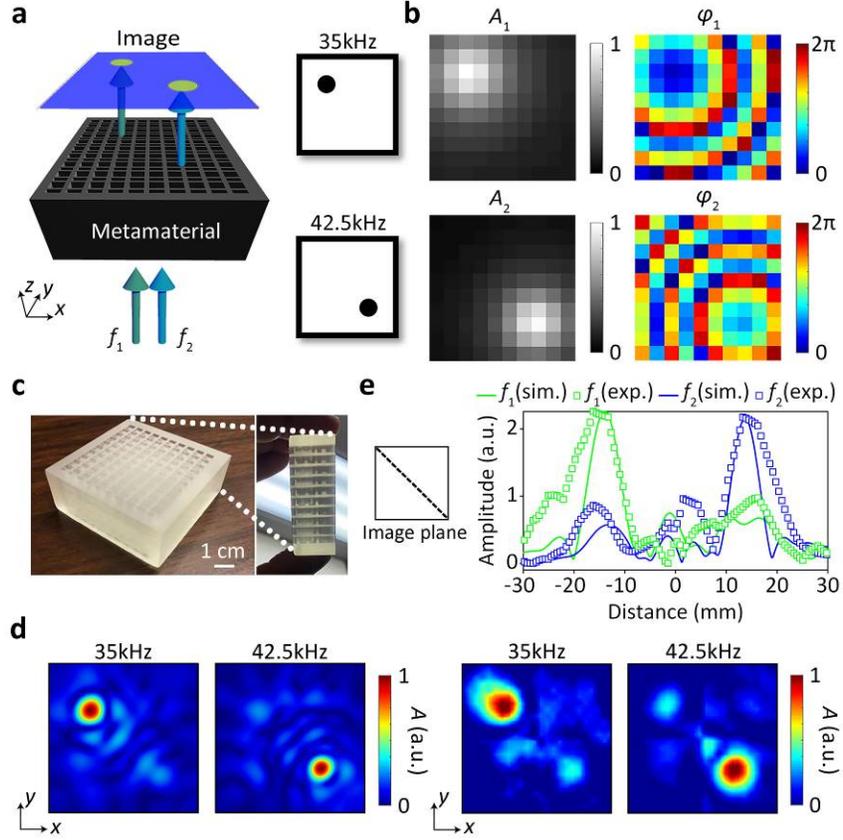

**Fig. 6** The design and demonstration of two-channel ultrasound lens. (**a**) The schematic diagram of the MAMH for two-channel ultrasound focusing. The image plane is 1 cm away ($z_l$ = 1 cm) from the hologram plane. The predesigned holography patterns are two focal points at the coordinate (-1.1 cm, 1.1 cm, 1 cm), and (1.1 cm, -1.1 cm, 1 cm), respectively. (**b**) Calculated amplitude and phase distributions for 35 kHz and 42.5 kHz, respectively. (**c**) The photograph of the 3-D printing sample, consisting of 10 × 10 unit cells with a size of 4.8 cm × 4.8 cm × 1.8 cm. Scale bar, 1 cm. (**d**) The simulated (left panel) and the measured (right panel) acoustic pressure amplitude distributions for 35 kHz and 42.5 kHz, respectively. Scale bar, 1 cm. (**e**) The simulated (sim., green and blue lines) and experimental (exp., green and blue squares) acoustic pressure amplitude distributions along the diagonal $x + y = 0$, $z = +1$ (shown in the inset). The horizontal axis is the distance between the targeted coordinate and the ordinate (0 cm, 0 cm, 1 cm).

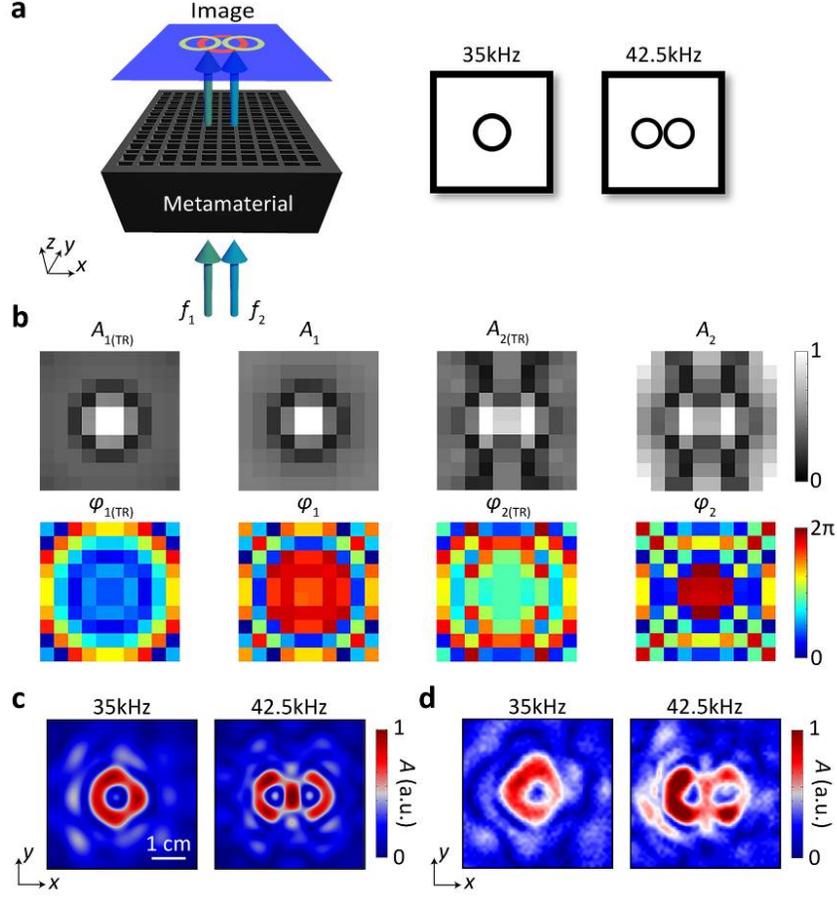

**Fig. 7** The design and demonstration of two-channel annular beams. (**a**) The schematic diagram of the MAMH for annular beams. The image plane is 1 cm away ($z_l$ = 1 cm) from the hologram plane. The predesigned holography pattern for $f_1$ = 35 kHz is one annular beam with the center coordinate (0 cm, 0 cm, 1 cm). The predesigned holography pattern for $f_1$ = 42.5 kHz is a pair annular beams with the center coordinates (-0.6 cm, 0 cm, 1 cm) and (0.6 cm, 0 cm, 1 cm). (**b**) The amplitude and phase distributions at hologram plane calculated by time reversal is ($A_{n(TR)}$, $\varphi_{n(TR)}$). The final amplitude and phase distributions of the sample are calculated as ($A_n$, $\varphi_n$). (**c**) The simulated acoustic pressure amplitude distributions of annular beam in *x-y* plane for 35 kHz and 42.5 kHz, respectively. Scale bar, 1 cm. (**d**) The measured acoustic pressure amplitude distributions of annular beam in *x-y* plane.

Additionally, to demonstrate highly precise wave reconstruction at the small wavelengths associated with 35 kHz and 42.5 kHz, we designed an ultrasonic MUMH that could generate distinctive annular beam patterns [40], that are dependent on the frequency of the incident wave. The schematic on Fig. 7(a) (left) illustrates the desired MUMH and shows that it would generate a single annular beam which has a center coordinate at (0 cm, 0 cm, 1 cm) when $f_1$ is incident, but when $f_2$ is incident, it would generate two annular beams that have two center coordinates at (-0.6 cm, 0 cm, 1 cm) and (0.6 cm, 0 cm, 1 cm). Figure 7(a) (right) shows the predesigned images that are taken into consideration in order to project these functionalities on to an image plane at $z_l=$ 1 cm. For this case, the amplitude and phase distributions that were calculated via time-reversal ($A_{n(\text{TR})}$, $\varphi_{n(\text{TR})}$), are shown in Fig. 7(b) for $f_1$ (left) and $f_2$ (right), respectively. The single air-coupled transducer was employed here as the source and was placed 3.5 cm behind the sample to ensure a good signal-to-noise ratio. However, since the incident wavefront here is not planar, amplitude and phase compensations were incorporated to the design, as done in the case of the audio meta-hologram, and the resultant ($A_n$, $\varphi_n$), for both frequencies, are shown adjacent to the initially calculated plots in Fig. 7(b). Figure 7(c) shows that the numerically predicted acoustic pressure amplitude distribution at the image plane, efficiently captures the details of the predesigned images for both $f_1$ and $f_2$. Figure 7(d) presents the experimentally measured pressure amplitude fields, that agree well with the numerical results and hence confirms that our ultrasonic meta-hologram generates distinctive channel/frequency dependent annular beams, whose centers lie at the desired coordinates. Such frequency dependent annular beam generation via a MUMH is highly desirable for acoustical tweezer-based particle manipulation [41].

**Discussion**

In summary, we designed and experimentally illustrated the working of multiplexed metamaterials for high fidelity acoustic holographic rendering. A new class of unit cells that could simultaneously modulate the amplitude and phase at two different frequencies were put forward and leveraged to build multiplexed meta-holograms that are encoded to project different predesigned images at different frequencies. The resultant meta-holograms not only introduce frequency as a new degree of freedom in metamaterial design, but also enable the generation of high-quality audio and ultrasonic wavefront reconstruction without the need for complex computational optimization. As a proof of concept, we numerically and experimentally demonstrated high-grade multiplexed audio holography, frequency dependent ultrasonic focusing, and two-channel annular beam generation. These demonstrated functionalities are benevolent for a wide-range of acoustic applications and could open avenues to next-generation devices for stereo-sound field reconstruction [37], ultrasonic therapy [39], and particle manipulation [41] based on switchable audio and ultrasonic devices.

**Methods**

**Numerical Simulations.** The simulations were performed using the commercial finite element analysis software, COMSOL Multiphysics 5.5a with the "Acoustic-Thermoviscous, Acoustic Interaction, Frequency Domain". The walls of the solid materials in the unit cells are set as sound hard boundaries. The considered mass density and the sound speed of background medium air are $\rho_0 = 1.21$ kg/m$^3$ and $c_0 = 343$ m/s, respectively.

**Sample Fabrications.** The sample for audio holography was fabricated using a commercial extrusion-based 3D printer (Ultimaker 3) and the material of poly-lactic acid (PLA). The samples for ultrasound experiments are fabricated using a commercial stereolithography 3D printer (Formlabs Form 3) and Formlabs Clear V4 resin. The 3D printed samples were rinsed in IPA in Form Wash for 10min and then post-cured by 405 nm light in Form Cure for 1hr.

**Experimental Measurements.** The experiments were carried out in a 3D space. For the measurement at 14 kHz and 17 kHz, a 10 cm-diameter loudspeaker was fixed at the incident side of the sample with a distance of 10 cm. The measurement of the acoustic intensity fields in Fig. 4 are obtained by scanning the output side of the sample with a distance of 5 cm (*viz.*, image plane), via 1/8-inch-diameter Brüel&Kjær type-4138-A-015 microphone and Brüel&Kjær PULSE Type 3160. The scanning area at image plane is 20 cm × 20 cm. For the measurement at 35 kHz and 42.5 kHz (the results for which are shown in Figs. 6 and 7), an air-coupled ultrasonic transducer (Murata MA40S4S) of diameter 1 cm-diameter was used as the source and placed at a distance, 3.5 cm behind the sample, for the annular beam measurement, and in the far field, for the focusing measurement. On the transmissive side, a GRAS Type 46BF free-field microphone of 1/4 inch diameter was translated by means of a linear scan stage over a square region of size, 4.8 cm × 4.8 cm. The data from the microphone was collected through a GRAS 12AA power module and Picoscope 4824 oscilloscope that helped improve the signal-to-noise ratio and offered a sufficiently high sampling frequency, respectively.

**Analytical derivation:** The amplitude and phase response of the unit cell can be analytically deduced by the transfer matrix method (See Supplementary Note 1)

**Acknowledgements**

This work is supported by the Air Force Office of Scientific Research under award number FA9550-18-1-7021, and by la Région Grand Est and Institut Carnot ICEEL. Y. J. would like to thank NSF for the support through CMMI 1951221. Work at LLNL was performed under the auspices of the U.S. Department of Energy by Lawrence Livermore National Laboratory under Contract DE-AC52-07NA27344.

**Author contributions**

Y.F.Z. and N.J.G. performed the theoretical simulations. Y.F.Z., L.Y.C., S.W.F. and B.A. conceived and designed the experiments for the audio meta-hologram. N.J.G., G.C.S. and Y.J. conceived and designed the experiments for the ultrasonic meta-holograms. X.X.X. and C.M.S fabricated the samples for ultrasound. Y.F.Z., N.J.G., Y.J. and B.A. wrote the manuscript. Y.J. and B.A. guided the research. All authors contributed to data analysis and discussions.

**Additional information**

Competing financial interests: The authors declare no competing financial interests.